\documentclass[twocolumn,letterpaper,aps,prl,superscriptaddress,showpacs,floatfix]{revtex4-1}

\usepackage{amsmath}
\usepackage{amssymb}
\usepackage[utf8]{inputenc}
\usepackage{units}
\usepackage{hepunits}
\usepackage{graphicx}	

\usepackage{xspace}	

\newcommand{\cms}{\sqrt{s}}
\newcommand{\cmsNN}{\sqrt{s_{\mathrm{NN}}}}

\newcommand{\nchg}{N_\mathrm{ch}}
\newcommand{\avgnchg}{\langle N_\mathrm{ch} \rangle}
\newcommand{\dd}{\mathrm{d}}
\newcommand{\dndeta}{\dd N_\mathrm{ch}/\dd\eta}

\newcommand{\dndetaZero}{\dd N_\mathrm{ch}/\dd \eta|_{\eta \approx 0}}
\newcommand{\dndetaOne}{\dd N_\mathrm{ch}/\dd \eta|_{|\eta| = 0.5}}

\newcommand{\Npp}{N_\mathrm{pp}}
\newcommand{\Nqp}{N_\mathrm{qp}}
\newcommand{\Nqpp}{N_\mathrm{qpp}}
\newcommand{\Nnpp}{N_\mathrm{npp}}

\newcommand{\ee}{\mathrm{e^- + e^+}}
\newcommand{\ep}{\mathrm{e^{\pm}+p}}
\begin{document}

%
\title{ Scaling properties of  the mean multiplicity and pseudorapidity density  \\ 
in  $e^{-}+e^{+}$, $e^{\pm}$+p,  p($\bar{\mathrm{p}}$)+p, p+A and A+A(B) collisions}

%
%
%
\author{ Roy~A.~Lacey}
\email[E-mail: ]{Roy.Lacey@Stonybrook.edu}
\affiliation{Department of Chemistry, 
Stony Brook University, 
Stony Brook, NY, 11794-3400, USA}
\affiliation{Dept. of Physics, 
Stony Brook University, 
Stony Brook, NY, 11794, USA}

\author{Peifeng Liu} 
\affiliation{Department of Chemistry, 
Stony Brook University, 
Stony Brook, NY, 11794-3400, USA}
\affiliation{Dept. of Physics, 
Stony Brook University, 
Stony Brook, NY, 11794, USA}
\author{ Niseem Magdy} 
\affiliation{Department of Chemistry, 
Stony Brook University, 
Stony Brook, NY, 11794-3400, USA}
\author{ M.Csan\'ad} 
\affiliation{Department of Chemistry, 
Stony Brook University, 
Stony Brook, NY, 11794-3400, USA}
\affiliation{Eotvos University, Department of Atomic Physics, H-1117 Budapest, Hungary}
\author{B. Schweid} 
\affiliation{Department of Chemistry, 
Stony Brook University, 
Stony Brook, NY, 11794-3400, USA}
\affiliation{Dept. of Physics, 
Stony Brook University, 
Stony Brook, NY, 11794, USA}
\author{ N.~N.~Ajitanand} 
\affiliation{Department of Chemistry, 
Stony Brook University, 
Stony Brook, NY, 11794-3400, USA}
\author{ J. Alexander} 
\affiliation{Department of Chemistry, 
Stony Brook University, 
Stony Brook, NY, 11794-3400, USA}
\author{ R. Pak}
\affiliation{Brookhaven National Laboratory, Upton, New York 11973, USA}

\date{\today}

\begin{abstract}

The pseudorapidity density ($\dndeta$) for p($\bar{\mathrm{p}}$)+p, p+A and A+A(B) collisions, 
and the mean multiplicity $\avgnchg$ for $\ee$,  $\ep$,  and p($\bar{\mathrm{p}}$)+p collisions, are 
studied for an inclusive range of beam energies ($\cms$). Characteristic scaling patterns  are 
observed for both $\dndeta$ and $\avgnchg$, consistent with a thermal  particle production mechanism for 
the bulk of the soft particles produced in all of these systems. They also validate an essential role for quark participants 
in these collisions. The scaled values for $\dndeta$ and $\avgnchg$ are observed to factorize into contributions 
which depend  on $\log(\cms)$ and  the number of nucleon or quark participant pairs $\Npp$. Quantification of these
contributions give  expressions which serve to systematize $\dndeta$ and $\avgnchg$  measurements spanning 
nearly four orders of magnitude in $\cms$,  and to predict their values as a function of $\cms$ and $\Npp$.

\end{abstract}

\pacs{25.75.Dw} 
	



\maketitle
 

%
Measurements of  particle yields and kinematic distributions in electron-positron ($\ee$), 
electron-proton ($\ep$),  proton-proton (p($\bar{p}$)+p), proton-nucleus (p+A) and 
nucleus-nucleus (A+A(B)) collisions,  are essential for characterizing the global properties 
of these collisions, and to develop a good understanding of 
the mehanism/s for particle production \cite{Kittel:2005fu,Abreu:2007kv,GrosseOetringhaus:2009kz,
Alver:2010ck,Aamodt:2010cz,Chatrchyan:2011pb,ATLAS:2011ag,Adamczyk:2012ku,Adare:2015bua}. 
The p+p measurements also provide crucial reference data for studies of nuclear-medium 
effects in A+A(B) and p+A collisions, as well as improved constraints to differentiate between particle 
production models and to fine-tune event generators.  

Particle production in A+A(B) collisions, is frequently described with 
thermodynamic and hydrodynamical models which utilize macroscopic variables 
such as temperature and entropy as model ingredients. 
This contrasts with the microscopic phenomenology  (involving ladders of perturbative gluons, classical 
random gauge fields or strings, and parton hadronization)  often used to characterize the soft collisions
which account for the bulk of the particles produced in $\ee$, $\ep$, p($\bar{\mathrm{p}}$)+p and p+A collisions
\cite{Kharzeev:2004if,Armesto:2004ud,Albrow:2006rt,Werner:2008zza,Dusling:2012wy}.  
The associated mechanisms, commonly classified as single-diffractive (SD) dissociation, 
double-diffractive (DD) dissociation and inelastic non-diffractive (ND) scattering in 
p($\bar{\mathrm{p}}$)+p collisions~\cite{Kittel:2005fu}, typically do not emphasize 
temperature and entropy as model elements.

Despite this predilection to use different theoretical model frameworks for p($\bar{\mathrm{p}}$)+p, p+A and A+A(B) collisions, 
it is well known that similar charged particle multiplicity ($\nchg$) and  pseudorapidity  density ($\dd \nchg/\dd\eta$)  
are obtained in p($\bar{\mathrm{p}}$)+p, and peripheral A+A(B) and p+A collisions. 
Moreover, an azimuthal long-range (pseudorapidity difference $|\Delta \eta| \geq 4$) 
two-particle angular correlation, akin to the ``ridge'' which results from collective anisotropic flow in A+A collisions, has  
been observed  in p+p and p+Pb collisions at the LHC \cite{Khachatryan:2010gv,Abelev:2012ola,Aad:2012gla,CMS:2012qk},
and in d+Au and He+Au collisions at RHIC \cite{Adare:2014keg,Adare:2015ctn}.  Qualitative consistency 
with these data has also been achieved in initial attempts  to describe the amplitudes of these 
correlations hydrodynamically \cite{Bozek:2011if,Adare:2014keg,Adare:2015ctn}. 
Thus, an important open question is whether equilibrium dynamics, linked to a common underlying particle 
production mechanism, dominates for these systems?

In this work, we use the available $\dndeta$ measurements for p+p, p+A and A+A(B) collisions, as well 
as the $\avgnchg$ measurements for $\ee$,  $\ep$, and  p($\bar{\mathrm{p}}$)+p collisions to search for scaling patterns 
which could signal such an underlying particle production mechanism.
%
%
\begin{figure*}[t] 
\includegraphics[width=0.85\linewidth]{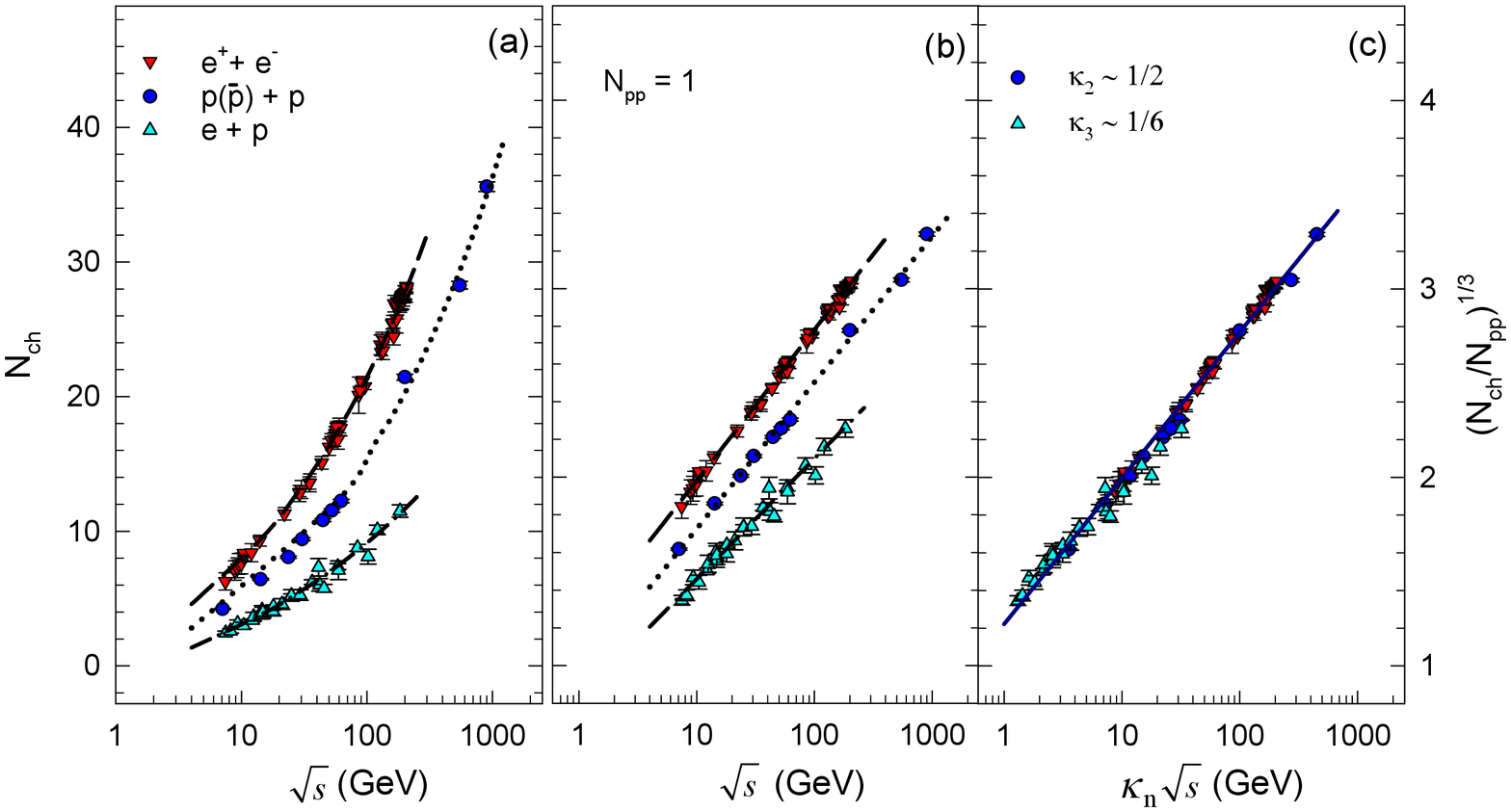}
\caption{(a) $\avgnchg$ vs. $\cms$ for $\ee$ \cite{Nakamura:2010zzi},  
p($\bar{\mathrm{p}}$)+p \cite{Benecke:1974if,Morse:1976tu,Breakstone:1983ns,Alner:1985wj,Ansorge:1988kn}
and  $\ep$ \cite{Adloff:1997fr,Breitweg:1999nt,Chekanov:2001sj} collisions; 
(b) $[\avgnchg/\Npp]^{1/3}$ vs. $\cms$; \\
(c)  $[\avgnchg/\Npp]^{1/3}$ vs. $\kappa_n \cms$ for the $\kappa_n$ values indicated.
The curves are drawn to guide the eye.
}
\label{Fig6}
\end{figure*}
%
%
%

Our scaling analysis employs the macroscopic entropy ($S$) ansatz
\begin{equation}
S\sim (TR)^3 \sim \mathrm{const.},
\label{eq1}
\end{equation}
to capture the underlying physics of particle production, where  $T$ is the temperature,  
$R$ is a characteristic size related to the volume, and  $\dndeta$ and $\avgnchg$ are
both  proportional to $S$.  A further simplification, $\Npp^{1/3} \propto R$, can be used to relate 
the number of participant pairs $\Npp$, to the initial volume. These pairs can be
specified as colliding participant pairs ({\em e.g.} $\Npp = 1$ for $\ee$, $\ep$ 
and p($\bar{\mathrm{p}}$)+p collisions), 
nucleon participant pairs ($\Nnpp$)  or quark participant pairs ($\Nqpp$). 
For p+p, p+A and A+A(B) collisions, Monte Carlo Glauber (MC-Glauber) 
calculations \cite{glauber,Lacey:2010hw,Eremin:2003qn,Bialas:2006kw,Nouicer:2006pr,Adler:2013aqf},
were performed for several collision centralities at each beam energy to obtain $\Nnpp$ and $\Nqpp$. 
In each of these calculations, a subset  $N_{\mathrm{np}} = 2N_{\mathrm{npp}}$ ($N_{\mathrm{qp}} = 2N_{\mathrm{qpp}}$) 
of the nucleons (quarks) become participants in each collision by undergoing an initial inelastic N+N (q+q) interaction.  
The N+N (q+q) cross sections used in these calculations were obtained from the data systematics reported in 
Ref.~\cite{Fagundes:2012rr}.

Equation~\ref{eq1} suggests similar characteristic patterns for 
$[(\dndeta)/\Npp]^{1/3}$ and $[\avgnchg/\Npp]^{1/3}$ as a function of centrality 
and $\cms$ for all collision systems. 
We use this scaling ansatz in conjunction with the wealth of measurements spanning 
several orders of magnitude in $\cms$, to search for, and study these predicted patterns.

%
%
\begin{figure*}[tb]
  \centering
  \begin{minipage}[b]{0.49\textwidth}
    \includegraphics[width=\textwidth]{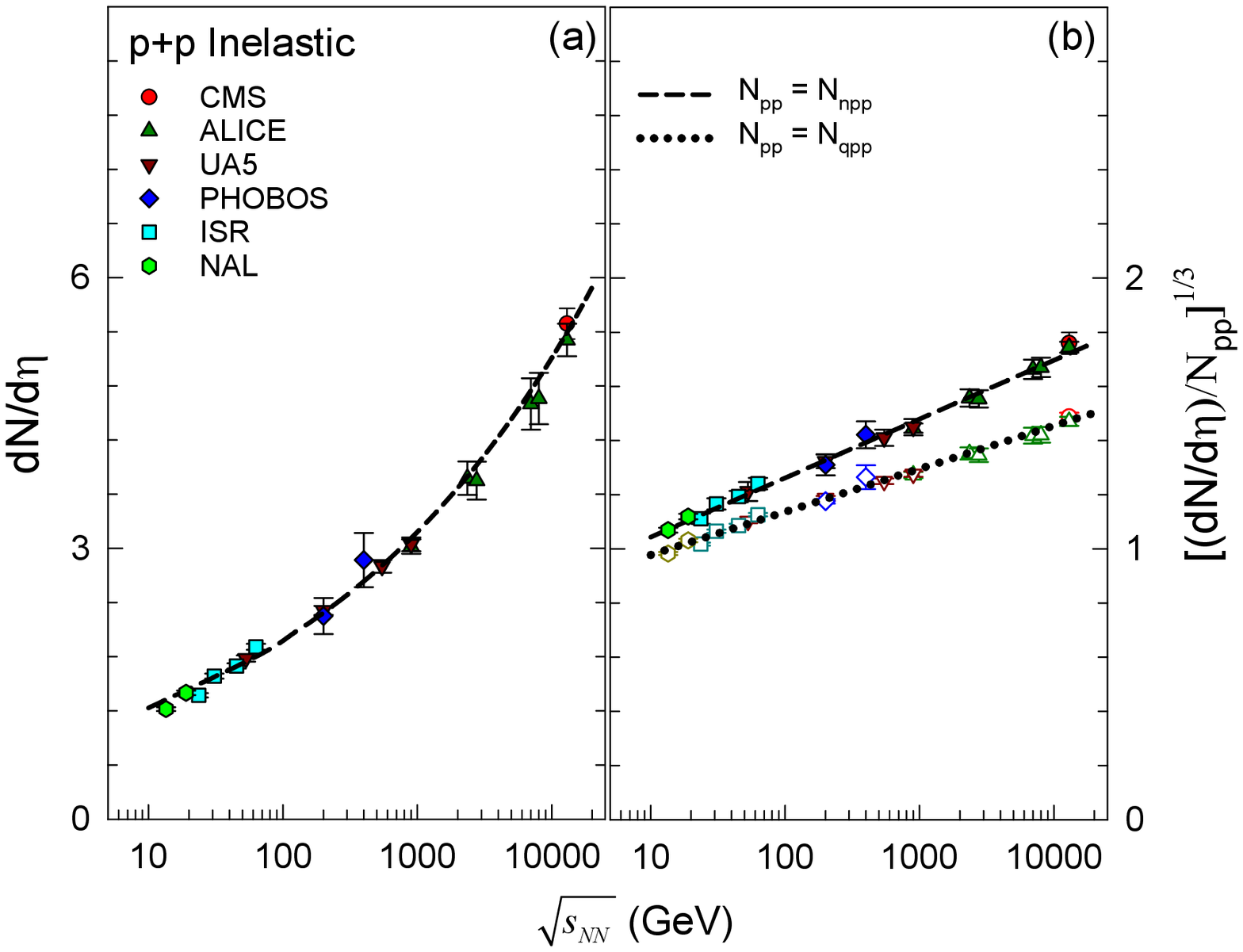}
   \caption{(a) $\dndetaZero$ vs. $\cmsNN$ and (b) $[(\dndeta)/\Npp]^{1/3}$ vs. $\cmsNN$, 
for p+p inelastic measurements from CMS \cite{Khachatryan:2015jna},  ALICE~\cite{Aamodt:2010ft,Adam:2015gka,Adam:2015pza}, 
UA5~\cite{Alner:1986xu}, PHOBOS~\cite{Nouicer:2004ke}, ISR~\cite{Breakstone:1983ns} and 
NAL Bubble Chamber~\cite{Whitmore:1973ri}. The error bars include systematic uncertainties, when available.
The curves in panel (b) represent fits to the data (see text).
}
\label{Fig1}
  \end{minipage}
  \hfill
  \begin{minipage}[b]{0.49\textwidth}
\includegraphics[width=\textwidth]{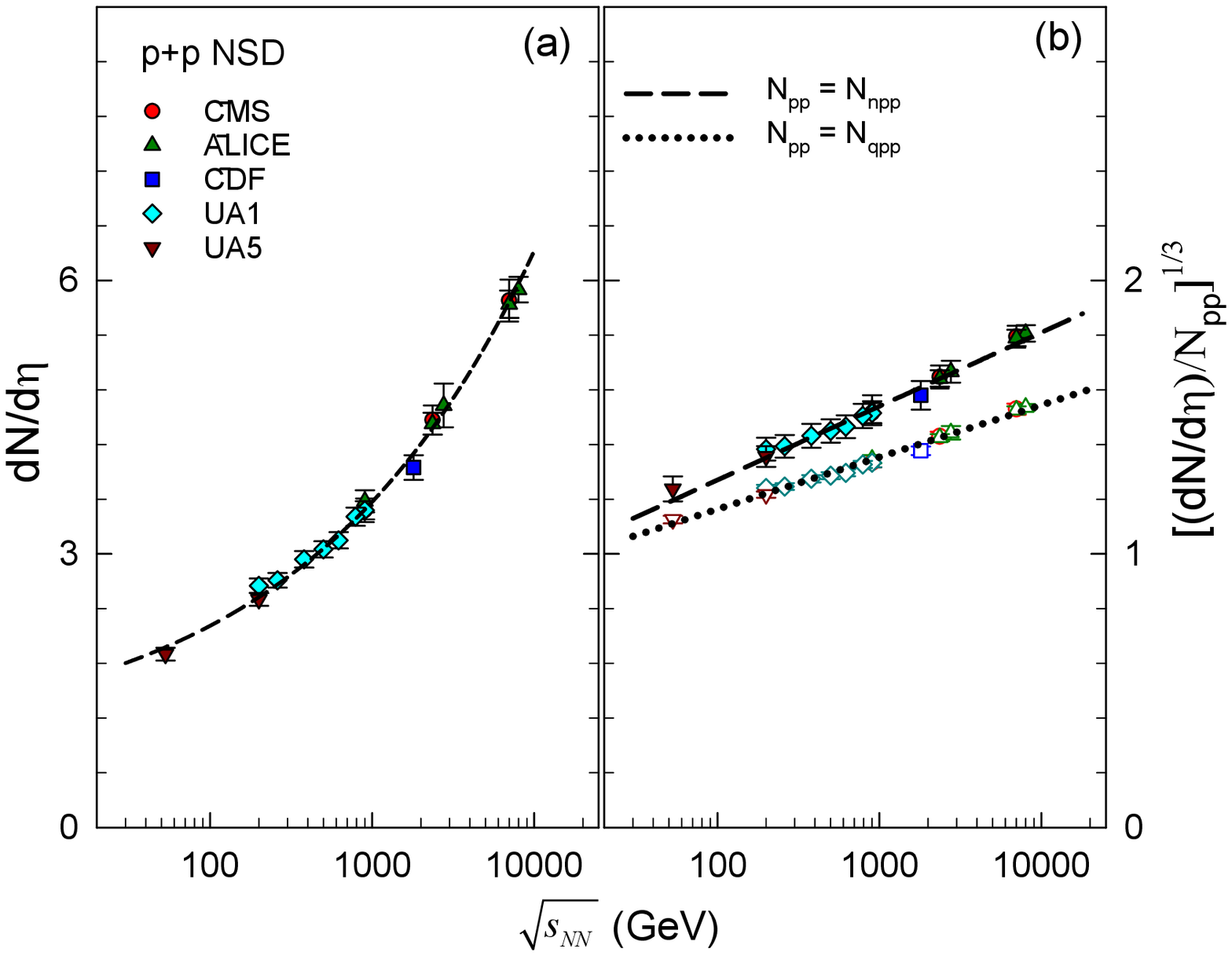}
\caption{(a) $\dndetaZero$ vs. $\cmsNN$ and (b) $[(\dndeta)/\Nnpp]^{1/3}$ vs. $\cmsNN$, 
for p+p NSD measurements from CMS \cite{Khachatryan:2010us},  ALICE~\cite{Adam:2015gka}, 
CDF~\cite{Abe:1989td}, UA1~\cite{Albajar:1989an} and UA5~\cite{Alner:1986xu}.
 The error bars include the available systematic uncertainties. The curves in panel (b) represent
fits to the data (see text).
}
\label{Fig2}
  \end{minipage}
\end{figure*}
%
%

%
%
\begin{figure*}[t] 
\vspace{+0.4cm}
\includegraphics[width=0.9\linewidth]{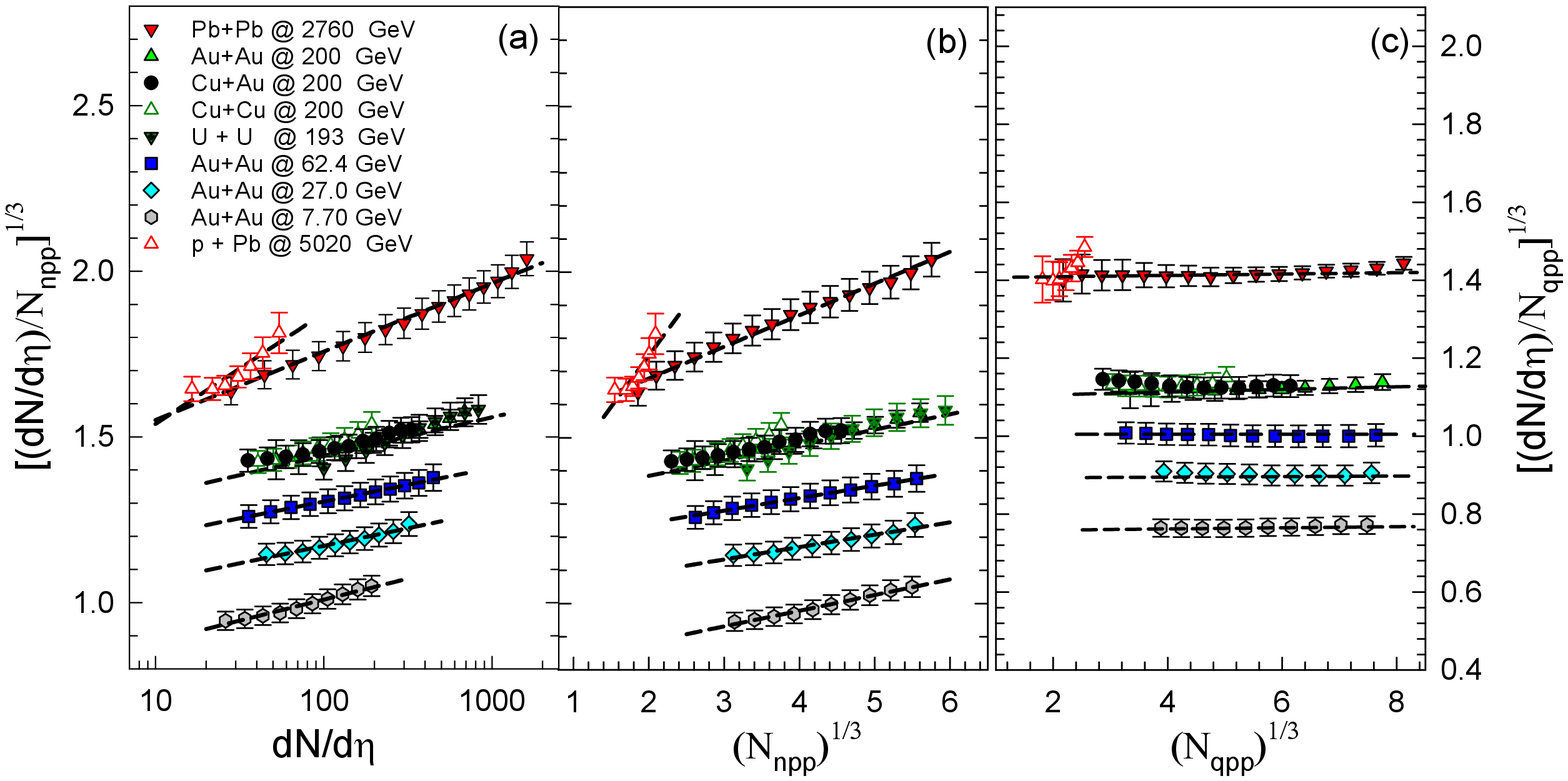}
\caption{(a) $[(\dndetaOne)/\Nnpp]^{1/3}$ vs. $\dndeta$; (b) $[(\dndetaOne)/\Nnpp]^{1/3}$ vs. $\Nnpp^{1/3}$;
(c) $[(\dndetaOne)/\Nqpp]^{1/3}$ vs. $\Nqpp^{1/3}$. Results are shown for several systems and $\cmsNN$ values as indicated. 
The data are obtained from Refs.~\cite{Alver:2010ck,Aamodt:2010cz,Chatrchyan:2011pb,ATLAS:2011ag,Aad:2015zza,Adamczyk:2012ku,Adare:2015bua}.
The curves are drawn to guide the eye.
}
\label{Fig3}
\end{figure*}

The $\avgnchg$ measurements for  $\ee$, $\ep$, and p($\bar{\mathrm{p}}$)+p collisions are shown in Fig.~\ref{Fig6}(a). 
They indicate a nonlinear increase with $\log(\cms)$, with 
$\avgnchg_{\mathrm{ee}} > \avgnchg_{\mathrm{pp}} > \avgnchg_{\mathrm{ep}}$ at each value of $\cms$. 
In contrast, Fig.~\ref{Fig6}(b) shows a linear increase of  $[\avgnchg/\Npp]^{1/3}$  ($\Npp =1$) 
with $\log(\cms)$, suggesting a linear increase of $T$ with $\log(\cms)$.  
Fig.~\ref{Fig6}(b) also indicates comparable slopes for  $[\avgnchg/\Npp]^{1/3}$ vs. $\log(\cms)$
for $\ee$, $\ep$, and p($\bar{\mathrm{p}}$)+p collisions, albeit with different magnitudes for $[\avgnchg/\Npp]^{1/3}$. 
This similarity is compatible with the notion of an effective energy $E_\mathrm{eff}$ in  
p($\bar{\mathrm{p}}$)+p and $\ep$ collisions, available for particle 
production \cite{Feinberg:1972ph,Albini:1975iu, Basile:1980ap,Basile:1981jz}. 
The remaining energy is associated with the leading particle/s which emerge 
at small angles with respect to the beam direction - the so-called leading particle effect \cite{Back:2003xk}. 
In a constituent quark picture~\cite{Nyiri:2002xb}, only a fraction of  the available quarks 
in p($\bar{\mathrm{p}}$)+p  and $\ep$ collisions,  contribute to $E_\mathrm{eff}$. 
Thus,  $\cms_{\mathrm{ee}} \approx \kappa_2\cms_{\mathrm{pp}} \approx \kappa_3\cms_{\mathrm{ep}}$ 
($\kappa_1 =1$)  would be expected to give similar values for $E_\mathrm{eff}$ \cite{Sarkisyan:2004vq} and  
hence, comparable $\avgnchg$ values in $\ee$,  p($\bar{\mathrm{p}}$)+p and $\ep$ collisions. 
Here, $\kappa_{2,3}$ are scale factors that are related to the number of quark participants and hence,
the fraction of the available c.m energy which contribute to particle production. 

Figure~\ref{Fig6}(c) validates this leading particle effect. It shows that the disparate magnitudes
of $[\avgnchg/\Npp]^{1/3}$ vs. $\cms$ for $\ee$, p($\bar{\mathrm{p}}$)+p and $\ep$ collisions (cf. Fig.~\ref{Fig6}(b))
scale to a single curve for $[\avgnchg/\Npp]^{1/3}$ vs. $\kappa_n \cms$
where $\kappa_1 = 1$, $\kappa_2 \sim 1/2$ and $\kappa_3 \sim 1/6$.
The values for $\kappa_{2,3}$ validate the important role of  quark participants
in p($\bar{\mathrm{p}}$)+p and $\ep$ collisions. A fit to the data in Fig.~\ref{Fig6}(c) gives  the expression
\begin{equation}
\begin{split}
\avgnchg = \left[ b_{\mathrm{\avgnchg}} + m_{\mathrm{\avgnchg}} \log(\kappa_n \cms) \right]^3, \\
b_{\mathrm{\avgnchg}} = 1.22 \pm 0.01,\  m_{\mathrm{\avgnchg}}= 0.775 \pm 0.006, 
\end{split}
\label{eq2}
\end{equation}
which can be used to predict $\avgnchg$ as a function of $\cms$, for $\ee$, $\ep$, and p($\bar{\mathrm{p}}$)+p collisions. 

Figures \ref{Fig1}(a) and  \ref{Fig2}(a) show $\dndetaZero$ measurements for inelastic (INE)
and non-single-diffractive (NSD)  p+p collisions (respectively) for beam energies spanning the range 
$\cmsNN \sim\unit{15}{\GeV}$~-~$ \unit{13}{\TeV}$; they indicate a monotonic 
increase of  $\dndetaZero$ with $\cmsNN$ similar to that observed for $\avgnchg$ in Fig.~\ref{Fig6}(a). 
Figs.~\ref{Fig1}(b) and  \ref{Fig2}(b), by contrast, confirms the expected linear growth of  $[(\dndetaZero)/\Nnpp]^{1/3}$ 
with $\log(\cmsNN)$. The open points and dotted curves in these figures, affirm 
the expected trend for the $\cms$ dependence of $[(\dndetaZero)/\Nqpp]^{1/3}$. Here, the change in magnitude largely
reflects the difference in the proportionality constants for $\Nnpp$ and $\Nqpp$ ({\em i.e.}, $\Npp^{1/3} \propto R$).
The fits, indicated by the dashed curves in Figs.~\ref{Fig1}(b) and  \ref{Fig2}(b), give the expressions
\begin{equation}
\begin{split}
\dndeta|_{\mathrm{INE}} =  \left[ b_{\mathrm{INE}} + m_{\mathrm{INE}} \log(\cmsNN)\right]^3, \\
b_{\mathrm{INE}} = 0.826 \pm 0.008,\  m_{\mathrm{INE}}= 0.220 \pm 0.004, 
\end{split}
\label{eq3}
\end{equation}
\begin{equation}
\begin{split}
\dndeta|_{\mathrm{NSD}} =  \left[ b_{\mathrm{NSD}} + m_{\mathrm{NSD}} \log(\cmsNN) \right]^3, \\
b_{\mathrm{NSD}} = 0.747 \pm 0.022,\   m_{\mathrm{NSD}}= 0.267 \pm 0.007,  
\end{split}
\label{eq4}
\end{equation}
for the mid-pseudorapidity density for INE and NSD p+p collisions.
Here, it is noteworthy that the recent inelastic p+p measurements  at $\cmsNN = \unit{13}{\TeV}$ 
by the CMS \cite{Khachatryan:2015jna} and  ALICE \cite{Adam:2015pza} collaborations  
are in very good agreement with the scaling prediction shown in Fig.~\ref{Fig1}(b). 
The data trends in Figs.~\ref{Fig6}(c), \ref{Fig1}(b) and \ref{Fig2}(b) also suggest 
that  the mean transverse momentum ($\left< p_T \right> \propto T$) for  the particles 
emitted in these collisions,  increase as $\log(\cms)$.

The scaling properties for p+A and A+A(B) collisions are summarized in  Fig.~\ref{Fig3} where illustrative 
plots of  $[(\dndetaOne)/\Nnpp]^{1/3}$ vs. $\dndeta$ and $\Nnpp^{1/3}$, 
and $[(\dndetaOne)/\Nqpp]^{1/3}$ vs. $\Nqpp^{1/3}$ are shown. 
Analogous plots were obtained for other collision systems and beam energies.
Figs.~\ref{Fig3}(a)  and \ref{Fig3}(b) show that, irrespective of the collision system, 
$[(\dndetaOne)/\Nnpp]^{1/3}$ increases as $\log(\dndeta)$ ($\Nnpp^{1/3}$),  
suggesting that $T$ has a logarithmic (linear) dependence on the pseudorapidity 
density (size) at a given  value of $\cmsNN$; note the slope increase with beam energy,
as well as the lack of sensitivity to system type (Cu+Cu, Cu+Au, Au+Au, U+U), for a fixed 
value of $\cmsNN$. These results suggest that, in addition to the expected increase with $\cmsNN$, 
the mean transverse momentum $\left< p_T \right>$ or transverse mass $\left< m_T \right>$ of the 
emitted particles, should increase as $\log(\dndeta)$ at a given value of $\cmsNN$.
They also suggest that the pseudorapidity density factorizes into 
contributions which depend on $\cmsNN$ and  $\Nnpp^{1/3}$ respectively. 
Indeed, the data sets shown for each $\cmsNN$ in Fig.~\ref{Fig3}(b), can be scaled 
to a single curve with scaling factors that are proportional to $\log(\cmsNN)$.

\begin{figure}[tb]
    \includegraphics[width=1.00\linewidth]{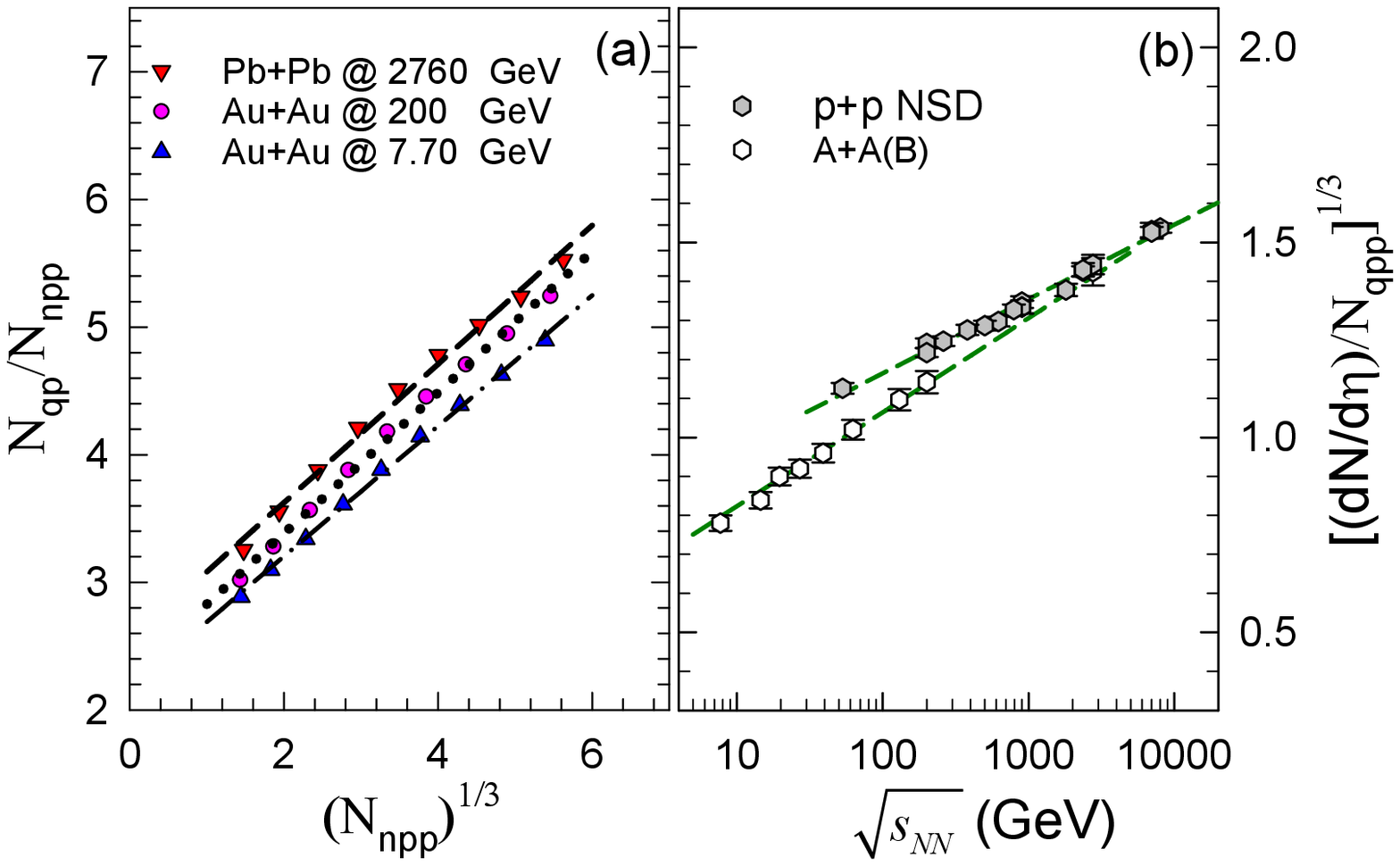}
   \caption{(a) $\Nqp/\Nnpp$ vs. $\Nnpp^{1/3}$ for Au+Au and Pb+Pb collisions; (b) $[(\dndetaOne)/N_{\mathrm{pp}}]^{1/3}$  vs. $\cmsNN$ for NSD p+p  
	and A+A(B) collisions as indicated. 
}
\label{Fig4}
\end{figure}

Figure~\ref{Fig3}(c)  contrasts with Figs.~\ref{Fig3}(a) and \ref{Fig3}(b). It shows that,
when $\Nqpp$ is used instead of $\Nnpp$,  the size dependence of $[(\dndetaOne)/\Nnpp]^{1/3}$, 
apparent in Fig.~\ref{Fig3}(b), is suppressed  (but not its $\cmsNN$ dependence).  
We attribute the flat dependence of $[(\dndetaOne)/\Nqpp]^{1/3}$ on size ($\Nnpp^{1/3}$ or $\Nqpp^{1/3}$), 
to the linear dependence of  $N_{\mathrm{qp}}/\Nnpp$ on initial size as illustrated in 
Fig.~\ref{Fig4}(a) for Pb+Pb and Au+Au collisions. Note that for central and mid-central 
p+Pb collisions, $N_{\mathrm{qp}}/\Nnpp$ decreases with $\Nnpp^{1/3}$; this 
results in a reduction of the energy deposited in these collisions, as well as large multiplicity 
fluctuations.

The $\cmsNN$ dependence of $[(\dndetaOne)/\Nqpp]^{1/3}$  for A+A(B) and  NSD p+p collisions  
are compared in Fig.~\ref{Fig4}(b). The comparison indicates strikingly similar trends  for NSD p+p, and A+A(B) 
collisions,  as would be expected for a common underlying particle production mechanism in these collisions. 
Note that for $\cmsNN \lesssim 2$~TeV, higher temperatures [and larger $\left< p_T \right>$] are implied for the 
smaller p+p collision systems. Figs.~\ref{Fig3}(c)  and  \ref{Fig4}(b) also indicate that the centrality and $\cms$ dependent 
values of $\dndetaOne$, obtained for different collision systems,  scale as $\Nqpp$ and $\log(\cmsNN)$. 
A fit to the A+A(B)  data in Fig.~\ref{Fig4}(b), gives the expression
\begin{eqnarray}
\begin{split}
\dndetaOne =   \Nqpp \left[ b_{\mathrm{AA}} + m_{\mathrm{AA}}\log(\cms) \right]^3 , \\
b_{\mathrm{AA}} = 0.530 \pm 0.008,\   m_{\mathrm{AA}}= 0.258 \pm 0.004, 
\end{split}
\label{eq5}
\end{eqnarray}
%
%
%
%
which systematizes the collision energy and centrality dependencies of the pseudorapidity density 
in A+A(B) collisions across the full range of beam energies.
Eq.~\ref{eq5} provides a basis for robust predictions of the value of $\dndetaOne$ as a function of $\Nqpp$ and $\cms$ 
across systems and collision energies. For example, it predicts an $\sim 20$\% increase in the $\dndetaOne$ values 
for Pb+Pb collisions (across centralities) at $\unit{5.02}{\TeV}$, compared to the same measurement 
at $\unit{2.76}{\TeV}$. This increase reflects the respective contributions linked to the increase   
in the value of  $\cms$ and the small growth in the magnitudes of $\Nqpp$.

%
%
%
%

%
In summary, we have performed a systematic study of the scaling properties of $\dndeta$ measurements for p+p, p+A and A+A(B) collisions, 
and $\avgnchg$ measurements for $\ee$,  $\ep$,  and p($\bar{\mathrm{p}}$)+p collisions, to investigate the 
mechanism for particle production in these collisions.
The wealth of the measurements, spanning several orders of magnitude in $\cms$, indicate characteristic scaling 
patterns  for both $\dndeta$ and $\avgnchg$, suggestive of a common underlying entropy production mechanism 
for these systems. The scaling patterns for $\avgnchg$ validate the essential role of the leading particle 
effect in p($\bar{\mathrm{p}}$)+p and $\ep$ collisions and the importance of quark participants in A+A(B) collisions. 
The patterns for the scaled values of $\dndeta$ and $\avgnchg$ indicate strikingly similar trends  for NSD p+p and 
A+A(B) collisions, and show that the pseudorapidity density and the $\avgnchg$ for $\ee$,  $\ep$, p+p, and 
A+A(B) collisions, factorize into contributions which depend on $\log(\cms)$ and  $\Npp$ respectively. 
The quantification of these scaling patterns, give expressions which serve to systematize the 
$\dndeta$ and $\avgnchg$  measurements for $\ee$,  $\ep$,  p($\bar{\mathrm{p}}$)+p, p+A  and A+A(B) collisions, 
and to predict their magnitudes as a function of $\Npp$ and $\cms$. These scaling results have important utility in the 
study of a broad array of observables which are currently being pursued at both RHIC and the LHC.

\section*{Acknowledgments}
This research is supported by the US DOE under contract DE-FG02-87ER40331.A008.

\bibliography{ref_mult_ms}   

\end{document}